\documentclass[aps,prl,twocolumn,]{revtex4-1}  

\usepackage{graphicx}  
\usepackage{dcolumn}   
\usepackage{bm}        
\usepackage{amssymb}   
\usepackage{color}

\usepackage{amsmath}

\usepackage{epstopdf}
\usepackage{bm}       
\usepackage{amsmath}
\usepackage{cancel}

\newcommand{\bee}{\begin{equation}}
\newcommand{\ee}{\end{equation}}
\newcommand{\beea}{\begin{eqnarray}}
\newcommand{\eea}{\end{eqnarray}}

\newcommand{\cO}{\ensuremath{\mathcal O} }
\newcommand{\X}{\ensuremath{\!\times\!} }
\newcommand{\lsim}{\ensuremath{\lesssim} }

\newcommand{\refcite}[1]{Ref.~\cite{#1}}
\newcommand{\eq}[1]{Eq.~\ref{#1}}
\newcommand{\fig}[1]{Fig.~\ref{#1}}

\begin{document}
\title{Finite size scaling of conformal theories in the presence of a near-marginal operator}

\author{Anqi~Cheng}
\affiliation{Department of Physics, University of Colorado, Boulder, CO 80309, USA}
\author{Anna~Hasenfratz}
\affiliation{Department of Physics, University of Colorado, Boulder, CO 80309, USA}
\author{Yuzhi~Liu}
\affiliation{Department of Physics, University of Colorado, Boulder, CO 80309, USA}
\author{Gregory Petropoulos}
\affiliation{Department of Physics, University of Colorado, Boulder, CO 80309, USA}
\author{David~Schaich}
\thanks{Present address: Department of Physics, Syracuse University, Syracuse, NY 13244, USA}
\affiliation{Department of Physics, University of Colorado, Boulder, CO 80309, USA}
\email{anna@eotvos.colorado.edu}

\begin{abstract}
The  slowly evolving gauge coupling of   gauge-fermion systems near the conformal window makes  numerical investigations of these models challenging. We consider finite size scaling and show that this  often used technique leads to inconsistent results if the leading order scaling corrections are neglected. When the corrections are included the results become consistent not only between different operators but  even when  data obtained at different gauge couplings or with different lattice actions are combined. Our results indicate that the SU(3) 12-fermion system is conformal with mass anomalous dimension $\gamma_m=0.235(15)$. 
\end{abstract}

\maketitle



Strongly coupled gauge-fermion systems near the conformal window are   candidates to describe the dynamics of electroweak symmetry breaking and beyond-Standard Model physics. 
These models are  expected to have a ``walking"  gauge coupling  and large anomalous mass dimension that can give rise to an enhanced fermion condensate,  while the weakly broken conformal symmetry could lead to a light dilaton that  plays the role of the Higgs boson~\cite{Appelquist:2013sia}. 
While the  non-perturbative properties of these systems are well suited to lattice studies,   standard lattice methods are frequently not efficient to investigate the infrared properties of near-conformal systems. 
The problems are mainly due to  the nearly marginal walking nature of the irrelevant gauge coupling. In this paper we investigate finite size scaling (FSS), a well established method to predict critical scaling exponents,
 and show that it is essential to take into account the effect of the nearly marginal  coupling in correction to scaling to obtain consistent results. 

We concentrate on the SU(3) gauge model with 12 fundamental fermions,
a controversial system.
Several groups have studied the infrared properties of this model using different methods and different lattice actions, arriving at contradictory conclusions regarding its infrared dynamics.
(For a limited set of references see Refs.~\cite{Appelquist:2009ty, Deuzeman:2009mh,  Fodor:2011tu, Appelquist:2011dp,DeGrand:2011cu, Hasenfratz:2011xn,Cheng:2011ic, Cheng:2013eu, Fodor:2012uw, Fodor:2012et,  Aoki:2012eq, Aoki:2013pca, Itou:2012qn, Lin:2012iw, Jin:2012dw}.) In particular FSS was considered in Refs.~\cite{Fodor:2011tu,Appelquist:2011dp,DeGrand:2011cu,Fodor:2012et,Aoki:2012eq}. Inconsistencies of the scaling exponent as predicted by different operators lead some authors to strongly question the conformal  behavior of this model.

We investigate this system at many gauge coupling values, and  also analyze the published meson spectrum data of the Lattice Higgs (LH) and LatKMI collaborations~\cite{Fodor:2011tu,Aoki:2012eq}. We develop a simple  formalism that  takes into account the effect of the leading irrelevant coupling and find consistent FSS  for several operators. The conclusion is further strengthened  when we combine several gauge couplings,  even  different lattice actions  together. These results suggest conformal infrared dynamics reinforcing the interpretation suggested by our earlier studies of the bare step scaling function~\cite{Hasenfratz:2011xn}, phase transitions~\cite{Hasenfratz:2013uha} and Dirac eigenvalues~\cite{Cheng:2013eu}.
 Preliminary results of our investigations have been reported in \refcite{Hasenfratz:2013eka}.

\begin{figure*}[tb]
\centering
  \includegraphics[width=0.45\linewidth]{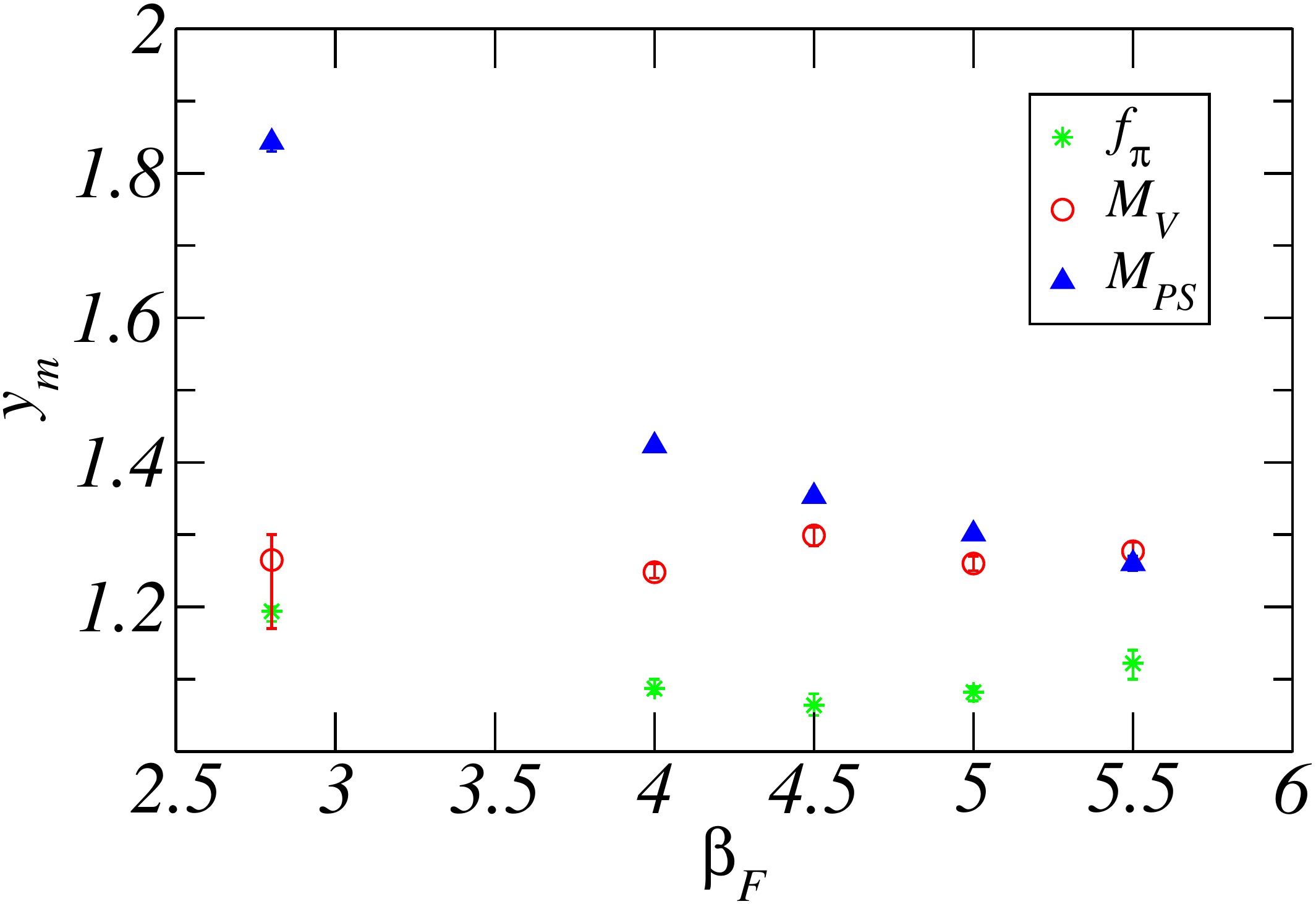} \hfill
  \includegraphics[width=0.45\linewidth]{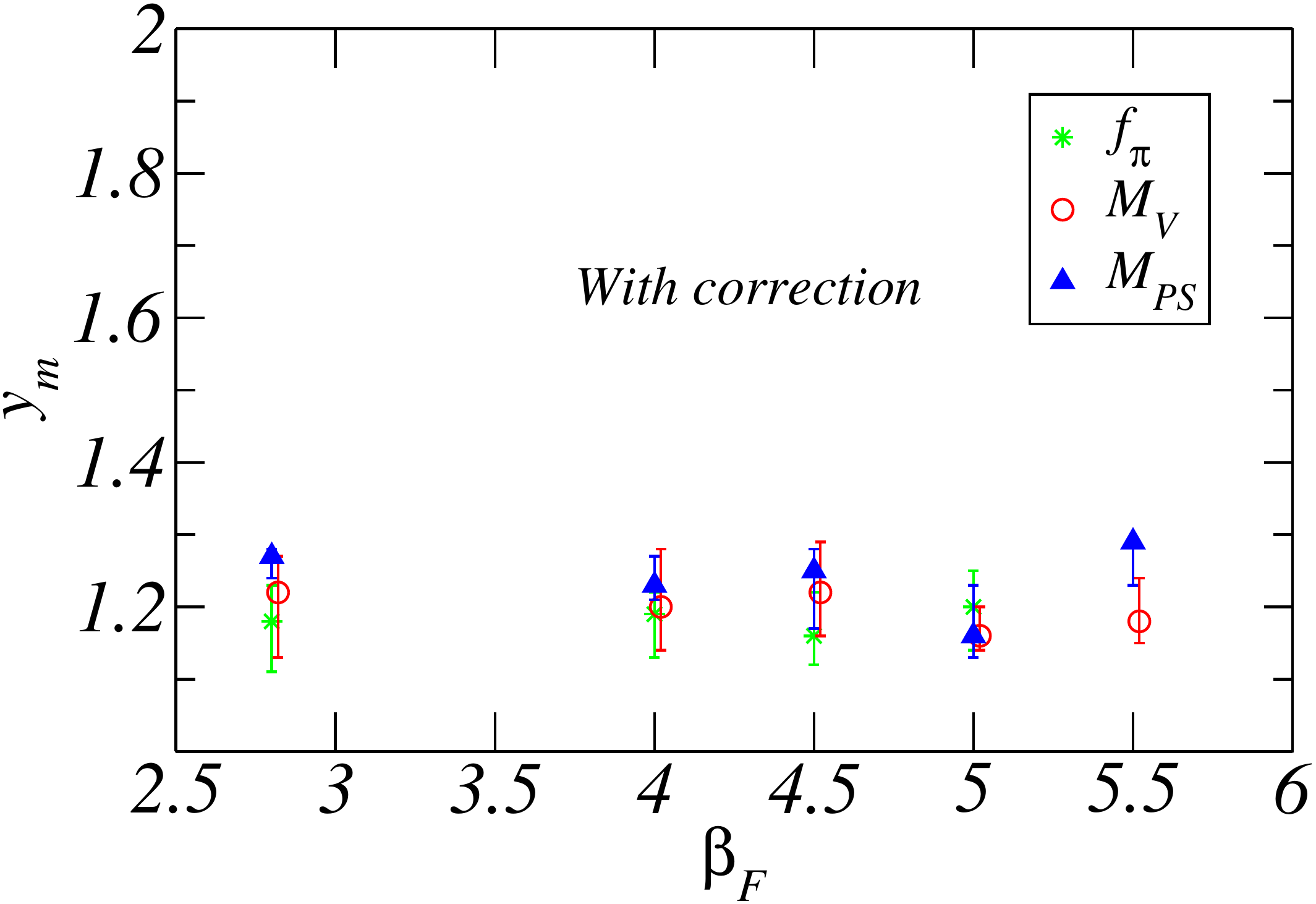}
   \caption{\label{fig:scaling_exp} The scaling dimension $y_m$ predicted by FSS, as a function of the gauge coupling $\beta_F$ for the $M_{PS}$ (blue triangles), $M_V$ (red circles) and $f_\pi$ (green $\times$s).  Left: fits including only the relevant mass operator (\protect\eq{eq:leading_fss}).  Right: fits including both the relevant operator and leading irrelevant corrections (\protect\eq{eq:corrected}) with $y_0=-0.36$ fixed at the two-loop value.}
 \end{figure*}

\section{Finite size scaling}

Finite size scaling is a well understood technique  to  investigate the critical properties of systems governed by one relevant operator.
Its derivation is easiest using renormalization group analysis and has been reviewed recently in connection with infrared conformal systems \cite{DeGrand:2009mt, DelDebbio:2010ze}.
Corrections to scaling due to the leading irrelevant operator have been successfully incorporated in  three dimensional spin model studies, for example in Refs.~\cite{Hasenbusch:1999mw,Hasenbusch:2011yya} that considered FSS of Ising model like systems on the critical surface.  A recent publication~\cite{DelDebbio:2013qta} investigated  corrections to scaling  in mass deformed conformal systems  in infinite volume.
In this work we consider FSS of mass deformed conformal systems including leading corrections.

For concreteness we consider a system with one relevant operator $m$ with scaling dimension $y_m > 0$.
All other operators, denoted by $g_i$, are irrelevant with scaling exponents $y_i < 0$.
Renormalization group arguments predict that in a finite spatial volume $L^3$, any physical quantity ``$M_H$'' with mass (engineering) dimension $[M_H] = 1$ depends only on specific combinations of the couplings, and can be written as
\begin{equation}
  \label{eq:general}
  M_H = L^{-1} f\left(x, g_i m^{-y_i / y_m}\right),
\end{equation}
where $x \equiv L m^{1 / y_m}$.
In the critical $m\to 0$ limit, $g_i m^{-y_i / y_m} \to 0$ and we find the familiar FSS formula
\begin{equation}
\label{eq:leading_fss}
  M_H = L^{-1} f_H(x),
\end{equation}
where $f_H(x)$ is an arbitrary but unique scaling function that
depends on the observable $M_H$. The exponent $y_m$  is universal and characteristic of the corresponding fixed point.

If one of the irrelevant operators, let's say $g_0$, is nearly marginal with scaling exponent $y_0 \lsim 0$, the term $g_0 m^{-y_0 / y_m}$ can remain significant and has to be included in the scaling analysis.
This leads to the modified FSS formula
\begin{equation}
  M_H = L^{-1} f_H\left(x, g_0 m^{\omega}\right),
\end{equation}
where $\omega \equiv -y_0 / y_m \gtrsim 0$.
The scaling function $f_H\left(x, g_0 m^{\omega}\right)$ is analytic even at the fixed point, and can be expanded as
\begin{equation}
  \label{eq:expansion}
  L M_H = F_H(x)\left\{1 + g_0 m^{\omega} G_H(x) + \cO\left(g_0^2 m^{2\omega}\right)\right\}.
\end{equation}
The first term is the usual FSS expression while the second  accounts for the leading corrections to scaling.

In the limit $x \to 0$, both $F_H(x)$ and $G_H(x)$ approach finite constants.
In the infinite-volume limit, with small but fixed $m$, $F_H(x)\propto x$ while $G_H(x)$ remains finite.
Our simulations cover a limited range $0.5 \lsim x \lsim 3$, over which we approximate $G_H(x)$ by a constant, $G_H(x) = c_G$, so 
\begin{equation}
  \label{eq:corrected}
  \frac{L M_H}{1 + c_G g_0 m^{\omega}} = F_H(x).
\end{equation}
One can test the validity of this approximation by  using only subsets of the data restricted to smaller ranges in $x$.
Equation \ref{eq:corrected} is very similar to the original \eq{eq:leading_fss}, 
however, the analysis now involves three parameters: $c_0 \equiv c_G g_0$, $y_0$ and $y_m$.

\section{Finite size scaling fits}
In our numerical studies we use nHYP smeared staggered fermions and a  gauge action that combines fundamental and adjoint plaquette terms with $\beta_A/\beta_F=-0.25$.
In \refcite{Cheng:2011ic} we reported on the phase structure and other properties of this action with $N_f=12$ fundamental fermions.

In the present work we consider gauge couplings 
$\beta_F=2.8$, 4.0, 4.5 and 5.0  and volumes $12^3\X24$, 
$16^3\X32$, $20^3\X40$, $24^3\X48$ and $32^3\X64$. The bare mass varies in the range $0.005 \leq m \leq 0.12$, such that the vector meson mass $a M_V < 0.7$.

In the FSS analysis we approximate $F_H(x)$ with two independent quadratic polynomials, one at $x<x_0$ and the other at $x > x_0$. We minimize the $\chi^2$ of this fit in terms of the polynomial coefficients, $x_0$, $y_m$, $y_0$ and $c_0$ using a Bayesian fitter based on \cite{Lepage:2001ym,Hornbostel:2011hu}\footnote{\texttt{https://github.com/gplepage/corrfitter}}. Priors on the values are 0.1 $\pm$ 20 for polynomial coefficients, 0.5 $\pm$ 20 for $-y_0$, 1.4 $\pm$ 1 for $y_m$, and -0.1 $\pm$ 5 for $c_0$. Table \ref{table:results} collects the results of several different fits, listing the relevant fit parameters as well as  $\chi^2$ per degrees of freedom (dof). This  $\chi^2$ represents not only the goodness of the FSS ``curve collapse" but the correctness of our rather simple fitting form for $F_H(x)$.
While the latter  could be improved by using a more elaborate fit function, we found the two independent quadratic polynomials to be sufficient.

Two loop perturbation theory predicts that the 12 flavor system is conformal with scaling exponent $y_m \approx 1.45$ and leading irrelevant exponent $y_0 \approx -0.36$. First we analyze the data  using the usual form of \eq{eq:leading_fss}, ignoring corrections to scaling. We consider each operator and $\beta_F$ data set independently.  The first row of  Table \ref{table:results} shows the result of the fit for the pseudoscalar mass $M_{PS}$  at $\beta_F=4.0$. This gauge coupling matches rather closely the published $\beta=2.2$ data of the LH collaboration and our prediction for  $y_m$ is  consistent with~\refcite{Fodor:2011tu}. 

The left panel of \fig{fig:scaling_exp} shows the results of similar analysis for the scaling exponent $y_m$  at other $\beta_F$ values for the pseudoscalar $M_{PS}$ and vector meson $M_V$ masses and $f_\pi$.
The scaling exponent shows significant variations between the three observables and as the function of $\beta_F$, suggesting that there is no consistent FSS when using the form of \eq{eq:leading_fss}.

When we take into account the leading scaling corrections according to \eq{eq:corrected} the situation changes. We are not able to constrain  the exponent $y_0$ using individual data sets so at this stage we fix $y_0=-0.36$, the perturbative  2-loop value. The correction term decreases  $\chi^2$ by more than a factor of two    as the second row of Table \ref{table:results} shows. We obtain consistent results when fitting only the small ($x<1.4$) or large ($x>1.1$) regions, justifying our approximation of constant $G(x) = c_G$. 

Repeating this analysis at other gauge couplings
leads to the results plotted on the right panel of \fig{fig:scaling_exp}, showing consistency between all three operators in the whole $\beta_F$ range investigated.
Not surprisingly the errors are significantly larger than before, especially for $f_\pi$ where the data constrain the correction coefficient $c_0$ only weakly.
\begin{table}[tdp]
\begin{center}
\begin{tabular}{|c|c|c|c|c|c|c|}
\hline
Op. & $\beta$ & $y_m$ & $y_0$ & $c_0$ (PS) &  $ s_m $  & $\chi^2$[dof]   \\
\hline\hline
PS	&	4.0	& 	1.421(3)	&	 - 		    &	0	        &	-	    	&	3.3[35]	\\
\hline
PS	&	4.0	&	1.223(17)	&	-0.36(fixed)	    &	-0.66(5)	&	-	    	&	1.3[36] \\
\hline
PS	&	4.0	& 	1.228(16)	&	 -0.499(58) &	-0.70(6)	&	1	    &   1.1[58] \\
	&	4.5	&		        &			    &	-0.50(6)	&	0.73   &	 	        \\
\hline
PS	&	2.8	& 	1.248(13)	&	 -0.466(16) &	-1.27(2)	&	3.03    &	2.9[99] \\
	&	4.0	& 	         	&	        	&	-0.60(4)	&	1	    &		    \\
	&	4.5	&		        &		        &	-0.40(5)	&	0.73	&	        \\
	&	5.0	&		        &		        &	-0.33(5)	&	0.58	&	        \\
\hline
PS	&	4.0	& 	1.238(13)	&	 -0.508(55) &	-0.67(5)	&	1	    &	1.4[95] \\
	&	4.5	&		        &		        &	-0.46(5)	&	0.73	&	        \\
	&	LH&&		&	-0.82(6)  &	1.11 	& 	 \\
	&	KMI 3.7&&	&	-0.76(6)	&	0.64    &	 	 \\
	&	KMI 4.0&& &	-0.70(5)	&	0.55    &	\\
\hline
PS,&	2.8	& 	1.228(11)	&	 -0.446(14) &	-1.28(2) 	&	3.03	    &	2.4[191] \\
V	&	4.0	& 	         	&	        	&	-0.66(3)   & 	1	    &		    \\
	&	4.5	&		        &		        &	-0.48(4) 	&	0.73	&	        \\
	&	5.0	&		        &		        &	-0.41(4) 	&	0.58	&	        \\
\hline
PS,&   2.8	&1.241(11)&-0.465(14)  &	-1.28(2)	    &	3.03	&   3.0[283] \\
V,	&	4.0	&		        &			    &	-0.62(3)	&	1       &	 	        \\
$f_\pi$&	4.5	&		        &			    &	-0.43(4)	&	0.73    &	 	        \\
	&	5.0	&		        &			    &	-0.36(4)	&	0.58    &	 	        \\
\hline
\end{tabular}

\end{center}
 \caption{Results of the FSS analysis in the 12 flavor system. $M_{PS}$,  $M_V$ and $f_\pi$ are analyzed at various $\beta_F$ couplings with the nHYP action, combined with the published data of the LH and LatKMI collaborations ~\protect\cite{Fodor:2011tu,Aoki:2012eq}.  
 $c_0$ denotes the amplitude of the leading correction (given only for the pseudoscalar)  and $s_m$ is the matching scale factor of the bare mass relative to the $\beta_F=4.0$ nHYP data. The last column lists the $\chi^2$ per degrees of freedom and dof of the fit.  }

\label{table:results}
\end{table}


\begin{figure*}[tbp]
\centering
  \includegraphics[width=0.45\linewidth]{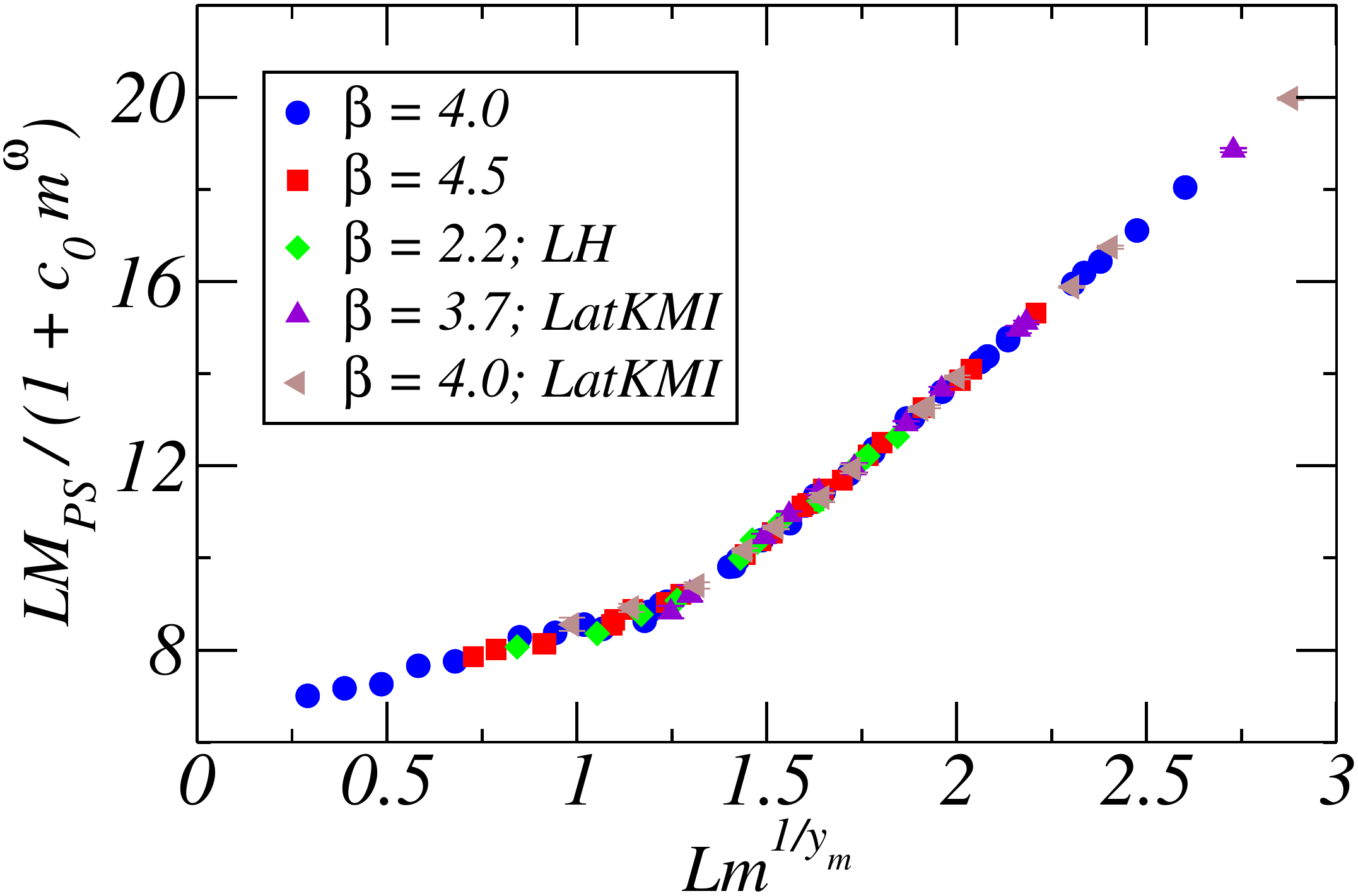}\hfill
  \includegraphics[width=0.45\linewidth]{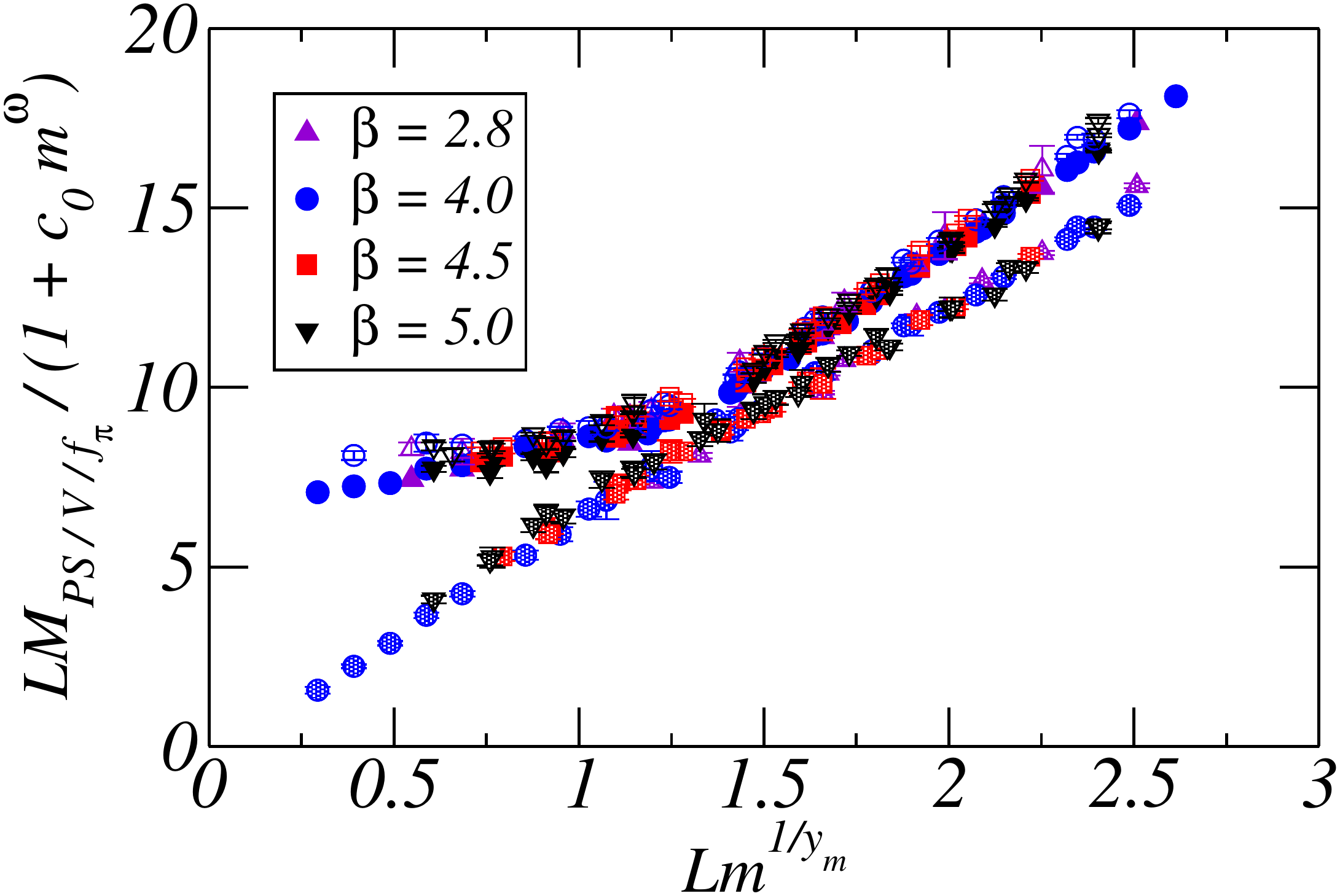}
  \caption{\label{fig:fss_combined} Left panel: The best curve collapse fit for the $M_{PS}$ combining  data at $\beta_F=4.0$, 4.5 and the published data of the LH and LatKMI collaborations ~\protect\cite{Fodor:2011tu,Aoki:2012eq} . The fit parameters are listed in Table~\protect\ref{table:results}. Right panel: Similar fit combining $M_{PS}$ (filled symbols), $M_V$ (open symbols), and $f_\pi$ (shaded symbols) using data at $\beta_F=2.8$, 4.0, 4.5 and 5.0.  The values for $f_\pi$ are rescaled by a factor of 9 for better clarity.}
\end{figure*}

If the gauge coupling is an irrelevant operator, the scaling function $F_H(x)$ is independent of $\beta_F$ and
we can significantly strengthen  the FSS fit   by combining data from different gauge couplings. This requires the introduction of a set of new parameters  $s_m$ that rescale the bare mass at each gauge coupling  $m \to s_m m$ to a common reference value. We choose $s_m=1$ at $\beta_F=4.0$. While the scale factors $s_m$ depend on the gauge coupling, they are independent of the operator. 
Such  global fits allow us to determine both  scaling dimensions  $y_m$ and $y_0$. However higher order corrections to scaling not accounted for in \eq{eq:corrected} can be different for different data sets and  significantly increase  $\chi^2$, especially when very different couplings are combined. 

The third entry in Table~\ref{table:results} shows the results of a fit to the pseudoscalar mass that combines couplings $\beta_F=4.0$ and 4.5. The universal fit parameters $y_m$ and $y_0$ are consistent with previous values with similar  $\chi^2$/dof. Including  the data sets at $\beta_F=2.8$ and 5.0 does not change the predicted values,  though we observe a significant increase in $\chi^2$ as the fourth entry of Table~\ref{table:results}  shows. This is not  surprising considering that the scale factors $s_m$ change by a factor of five in this case. While the  scale factors  increase with decreasing $\beta_F$,  the $c_0$ coefficients decrease, suggesting that the conformal infrared fixed point (or its projection to the $\beta_F$ axis) where $c_0 = 0$ occurs at weaker gauge coupling. 

We can  combine  different lattice actions, not only  gauge couplings,  in the FSS fit. Both the LH and LatKMI collaborations ~\protect\cite{Fodor:2011tu,Aoki:2012eq} published some of their spectrum results which we can fit together with our nHYP action data. As the  next entry of Table~\ref{table:results} shows, such a combined fit  for $M_{PS}$ has a  small $\chi^2$/dof  with scaling dimensions consistent with previous fits.  The left panel of \fig{fig:fss_combined} illustrates the ``curve collapse" of this fit. 

The vector meson mass can be analyzed similarly,  and one can even combine it with the pseudoscalar. The fit now depends on
two independent scaling functions $F_H(x)$ for the two operators.
As the fifth entry in Table~\ref{table:results}  shows  the scaling dimensions from such a combined fit are consistent with the pseudoscalar fit  with similar $\chi^2$/dof. 
The  errors listed are statistical only and do not take into account possible correlations between the two operators.

The pion decay constant   is expected to scale with the same universal exponents as the hadron masses but it exhibits very different finite volume effects~\cite{Fodor:2011tu,Aoki:2012eq} and  corrections to  scaling could be more significant to $f_\pi$ than to the masses, especially in small volumes. As the volume decreases the mesonic bound states are squeezed.  At some point the volume might become too small to support bound states and the physical meaning of $f_\pi$ changes. The first order phase transition observed with Wilson fermions in \refcite{Ishikawa:2013tua} could be of similar origin.   Nevertheless a combined fit for $M_{PS}$, $M_V$, and $f_\pi$ is reasonable even when we combine the  gauge couplings $\beta_F=2.8$, 4.0, 4.5 and 5.0 as shown in the last entry of Table~\ref{table:results}.
In this case we fit nearly 300 data points with three independent scaling functions and 12 $c_0$ coefficients describing the leading corrections to scaling, yet the predicted scaling exponents are consistent with all previous fit results.

\section{Conclusion}
We have demonstrated that apparent inconsistencies in finite size scaling analysis of the $N_f=12$ system can be resolved by considering the effect of the leading irrelevant  coupling, at least for the pseudoscalar and vector meson masses and $f_\pi$.
For these quantities combined fits of several independent data sets  at different gauge couplings and even different lattice actions are consistent with conformal infrared dynamics.  Based on  various fits presented in Table~\ref{table:results} we predict the anomalous mass dimension $\gamma_m=y_m-1=0.235(15)$ at the corresponding infrared fixed point.

We have investigated only three physical quantities and cannot prove that all other observables will scale consistently once corrections to scaling are taken into account -- especially because these corrections might be more important to some observables than to others. 
It will  be important to consider other quantities, especially  those related to the static potential as published large volume data appear to be inconsistent with conformal dynamics~\cite{Fodor:2012uw}. 

We expect that systems near the conformal boundary will generically possess a nearly-marginal operator due to the walking gauge coupling, or possibly even some other operator that becomes relevant  at the conformal boundary. 
The results presented in this paper suggest that such an operator has important effects that have to be considered when studying  any strongly-coupled many-flavor system.


\section*{Acknowledgments} 
We thank  Biagio Lucini for suggesting that we include corrections in the FSS analysis, Roman Zwicky for useful discussions concerning the correction terms and Julius Kuti for his helpful probing questions.
Part of this work was performed when A.~H.\ and D.~S.\ visited the Aspen Center for Physics (NSF Grant No.~1066293) and CP$^3$-Origins in Odense, and we thank both institutions for their support and hospitality. A.~H. is grateful for the hospitality of the Brookhaven National Laboratory HET group during her extended visit.
This research was partially supported by the U.S.~Department of Energy (DOE) through Grant No.~DE-SC0010005 (A.~C., A.~H., Y.~L., and D.~S.) and by the DOE Office of Science Graduate Fellowship Program under Contract No.~DE-AC05-06OR23100 (G.~P.).
Our code is based in part on the MILC Collaboration's public lattice gauge theory software\footnote{\texttt{http://www.physics.utah.edu/$\sim$detar/milc/}}.
Numerical calculations were carried out on the HEP-TH and Janus clusters, partially funded by NSF Grant No.~CNS-0821794  at the University of Colorado; at Fermilab under the auspices of USQCD supported by the DOE; and at the San Diego Computing Center through XSEDE supported by National Science Foundation Grant No.~OCI-1053575.

{\renewcommand{\baselinestretch}{0.86} 

\begin{thebibliography}{27}%
\makeatletter
\providecommand \@ifxundefined [1]{%
 \@ifx{#1\undefined}
}%
\providecommand \@ifnum [1]{%
 \ifnum #1\expandafter \@firstoftwo
 \else \expandafter \@secondoftwo
 \fi
}%
\providecommand \@ifx [1]{%
 \ifx #1\expandafter \@firstoftwo
 \else \expandafter \@secondoftwo
 \fi
}%
\providecommand \natexlab [1]{#1}%
\providecommand \enquote  [1]{``#1''}%
\providecommand \bibnamefont  [1]{#1}%
\providecommand \bibfnamefont [1]{#1}%
\providecommand \citenamefont [1]{#1}%
\providecommand \href@noop [0]{\@secondoftwo}%
\providecommand \href [0]{\begingroup \@sanitize@url \@href}%
\providecommand \@href[1]{\@@startlink{#1}\@@href}%
\providecommand \@@href[1]{\endgroup#1\@@endlink}%
\providecommand \@sanitize@url [0]{\catcode `\\12\catcode `\$12\catcode
  `\&12\catcode `\#12\catcode `\^12\catcode `\_12\catcode `\%12\relax}%
\providecommand \@@startlink[1]{}%
\providecommand \@@endlink[0]{}%
\providecommand \url  [0]{\begingroup\@sanitize@url \@url }%
\providecommand \@url [1]{\endgroup\@href {#1}{\urlprefix }}%
\providecommand \urlprefix  [0]{URL }%
\providecommand \Eprint [0]{\href }%
\@ifxundefined \urlstyle {%
  \providecommand \doi  [0]{\begingroup \@sanitize@url \@doi}%
  \providecommand \@doi [1]{\endgroup \@@startlink {\doibase
  #1}doi:\discretionary {}{}{}#1\@@endlink }%
}{%
  \providecommand \doi  [0]{doi:\discretionary{}{}{}\begingroup
  \urlstyle{rm}\Url }%
}%
\providecommand \doibase [0]{http://dx.doi.org/}%
\providecommand \Doi [0]{\begingroup \@sanitize@url \@Doi }%
\providecommand \@Doi  [1]{\endgroup\@@startlink{\doibase#1}\@@Doi}%
\providecommand \@@Doi [1]{#1\@@endlink}%
\providecommand \selectlanguage [0]{\@gobble}%
\providecommand \bibinfo  [0]{\@secondoftwo}%
\providecommand \bibfield  [0]{\@secondoftwo}%
\providecommand \translation [1]{[#1]}%
\providecommand \BibitemOpen [0]{}%
\providecommand \bibitemStop [0]{}%
\providecommand \bibitemNoStop [0]{.\EOS\space}%
\providecommand \EOS [0]{\spacefactor3000\relax}%
\providecommand \BibitemShut  [1]{\csname bibitem#1\endcsname}%
\bibitem[{\citenamefont{Appelquist
  et~al.}(2013)\citenamefont{Appelquist, Brower, Catterall,
  Fleming, Giedt, Hasenfratz, Kuti, Neil, and Schaich}}]{Appelquist:2013sia}
  For brief reviews, see
\bibinfo{author}{\bibfnamefont{T.}~\bibnamefont{Appelquist}},
  \bibinfo{author}{\bibfnamefont{R.}~\bibnamefont{Brower}},
  \bibinfo{author}{\bibfnamefont{S.}~\bibnamefont{Catterall}},
  \bibinfo{author}{\bibfnamefont{G.}~\bibnamefont{Fleming}},
  \bibinfo{author}{\bibfnamefont{J.}~\bibnamefont{Giedt}},
  \bibinfo{author}{\bibfnamefont{A.}~\bibnamefont{Hasenfratz}},
  \bibinfo{author}{\bibfnamefont{J.}~\bibnamefont{Kuti}},
  \bibinfo{author}{\bibfnamefont{E.}~\bibnamefont{Neil}} \bibnamefont{and}
  \bibinfo{author}{\bibfnamefont{D.}~\bibnamefont{Schaich}}
  (\bibinfo{year}{2013}), \eprint{1309.1206},
  \urlprefix\url{http://www.usqcd.org/documents/bsm.pdf}, and references therein.
\bibitem [{\citenamefont {Appelquist}\ \emph {et~al.}(2009)\citenamefont
  {Appelquist}, \citenamefont {Fleming},\ and\ \citenamefont
  {Neil}}]{Appelquist:2009ty}%
  \BibitemOpen
  \bibfield  {author} {\bibinfo {author} {\bibfnamefont {T.}~\bibnamefont
  {Appelquist}}, \bibinfo {author} {\bibfnamefont {G.~T.}\ \bibnamefont
  {Fleming}}, \ and\ \bibinfo {author} {\bibfnamefont {E.~T.}\ \bibnamefont
  {Neil}},\ }\Doi {10.1103/PhysRevD.79.076010} {\bibfield  {journal} {\bibinfo
  {journal} {Phys. Rev.},\ }\textbf {\bibinfo {volume} {D79}},\ \bibinfo
  {pages} {076010} (\bibinfo {year} {2009})},\ \Eprint
  {http://arxiv.org/abs/0901.3766} {arXiv:0901.3766} \BibitemShut {NoStop}%
\bibitem [{\citenamefont {Deuzeman}\ \emph {et~al.}(2010)\citenamefont
  {Deuzeman}, \citenamefont {Lombardo},\ and\ \citenamefont
  {Pallante}}]{Deuzeman:2009mh}%
  \BibitemOpen
  \bibfield  {author} {\bibinfo {author} {\bibfnamefont {A.}~\bibnamefont
  {Deuzeman}}, \bibinfo {author} {\bibfnamefont {M.~P.}\ \bibnamefont
  {Lombardo}}, \ and\ \bibinfo {author} {\bibfnamefont {E.}~\bibnamefont
  {Pallante}},\ }\Doi {10.1103/PhysRevD.82.074503} {\bibfield  {journal}
  {\bibinfo  {journal} {Phys. Rev.},\ }\textbf {\bibinfo {volume} {D82}},\
  \bibinfo {pages} {074503} (\bibinfo {year} {2010})},\ \Eprint
  {http://arxiv.org/abs/0904.4662} {arXiv:0904.4662} \BibitemShut {NoStop}%
\bibitem [{\citenamefont {Fodor}\ \emph {et~al.}(2011)\citenamefont {Fodor},
  \citenamefont {Holland}, \citenamefont {Kuti}, \citenamefont {Nogradi},\ and\
  \citenamefont {Schroeder}}]{Fodor:2011tu}%
  \BibitemOpen
  \bibfield  {author} {\bibinfo {author} {\bibfnamefont {Z.}~\bibnamefont
  {Fodor}}, \bibinfo {author} {\bibfnamefont {K.}~\bibnamefont {Holland}},
  \bibinfo {author} {\bibfnamefont {J.}~\bibnamefont {Kuti}}, \bibinfo {author}
  {\bibfnamefont {D.}~\bibnamefont {Nogradi}}, \ and\ \bibinfo {author}
  {\bibfnamefont {C.}~\bibnamefont {Schroeder}},\ }\Doi
  {10.1016/j.physletb.2011.07.037} {\bibfield  {journal} {\bibinfo  {journal}
  {Phys. Lett.},\ }\textbf {\bibinfo {volume} {B703}},\ \bibinfo {pages} {348}
  (\bibinfo {year} {2011})},\ \Eprint {http://arxiv.org/abs/1104.3124}
  {arXiv:1104.3124} \BibitemShut {NoStop}%
\bibitem [{\citenamefont {Appelquist}\ \emph {et~al.}(2011)\citenamefont
  {Appelquist}, \citenamefont {Fleming}, \citenamefont {Lin}, \citenamefont
  {Neil},\ and\ \citenamefont {Schaich}}]{Appelquist:2011dp}%
  \BibitemOpen
  \bibfield  {author} {\bibinfo {author} {\bibfnamefont {T.}~\bibnamefont
  {Appelquist}}, \bibinfo {author} {\bibfnamefont {G.~T.}\ \bibnamefont
  {Fleming}}, \bibinfo {author} {\bibfnamefont {M.~F.}\ \bibnamefont {Lin}},
  \bibinfo {author} {\bibfnamefont {E.~T.}\ \bibnamefont {Neil}}, \ and\
  \bibinfo {author} {\bibfnamefont {D.}~\bibnamefont {Schaich}},\ }\Doi
  {10.1103/PhysRevD.84.054501} {\bibfield  {journal} {\bibinfo  {journal}
  {Phys. Rev.},\ }\textbf {\bibinfo {volume} {D84}},\ \bibinfo {pages} {054501}
  (\bibinfo {year} {2011})},\ \Eprint {http://arxiv.org/abs/1106.2148}
  {arXiv:1106.2148} \BibitemShut {NoStop}%
\bibitem [{\citenamefont {DeGrand}(2011)}]{DeGrand:2011cu}%
  \BibitemOpen
  \bibfield  {author} {\bibinfo {author} {\bibfnamefont {T.}~\bibnamefont
  {DeGrand}},\ }\Doi {10.1103/PhysRevD.84.116901} {\bibfield  {journal}
  {\bibinfo  {journal} {Phys. Rev.},\ }\textbf {\bibinfo {volume} {D84}},\
  \bibinfo {pages} {116901} (\bibinfo {year} {2011})},\ \Eprint
  {http://arxiv.org/abs/1109.1237} {arXiv:1109.1237} \BibitemShut {NoStop}%
\bibitem [{\citenamefont {Hasenfratz}(2012)}]{Hasenfratz:2011xn}%
  \BibitemOpen
  \bibfield  {author} {\bibinfo {author} {\bibfnamefont {A.}~\bibnamefont
  {Hasenfratz}},\ }\Doi {10.1103/PhysRevLett.108.061601} {\bibfield  {journal}
  {\bibinfo  {journal} {Phys. Rev. Lett.},\ }\textbf {\bibinfo {volume}
  {108}},\ \bibinfo {pages} {061601} (\bibinfo {year} {2012})},\ \Eprint
  {http://arxiv.org/abs/1106.5293} {arXiv:1106.5293} \BibitemShut {NoStop}%
\bibitem [{\citenamefont {Cheng}\ \emph {et~al.}(2012)\citenamefont {Cheng},
  \citenamefont {Hasenfratz},\ and\ \citenamefont {Schaich}}]{Cheng:2011ic}%
  \BibitemOpen
  \bibfield  {author} {\bibinfo {author} {\bibfnamefont {A.}~\bibnamefont
  {Cheng}}, \bibinfo {author} {\bibfnamefont {A.}~\bibnamefont {Hasenfratz}}, \
  and\ \bibinfo {author} {\bibfnamefont {D.}~\bibnamefont {Schaich}},\ }\Doi
  {10.1103/PhysRevD.85.094509} {\bibfield  {journal} {\bibinfo  {journal}
  {Phys. Rev.},\ }\textbf {\bibinfo {volume} {D85}},\ \bibinfo {pages} {094509}
  (\bibinfo {year} {2012})},\ \Eprint {http://arxiv.org/abs/1111.2317}
  {arXiv:1111.2317} \BibitemShut {NoStop}%
\bibitem [{\citenamefont {Cheng}\ \emph {et~al.}(2013)\citenamefont {Cheng},
  \citenamefont {Hasenfratz}, \citenamefont {Petropoulos},\ and\ \citenamefont
  {Schaich}}]{Cheng:2013eu}%
  \BibitemOpen
  \bibfield  {author} {\bibinfo {author} {\bibfnamefont {A.}~\bibnamefont
  {Cheng}}, \bibinfo {author} {\bibfnamefont {A.}~\bibnamefont {Hasenfratz}},
  \bibinfo {author} {\bibfnamefont {G.}~\bibnamefont {Petropoulos}}, \ and\
  \bibinfo {author} {\bibfnamefont {D.}~\bibnamefont {Schaich}},\ }\Doi
  {10.1007/JHEP07(2013)061} {\bibfield  {journal} {\bibinfo  {journal} {JHEP},\
  }\textbf {\bibinfo {volume} {1307}},\ \bibinfo {pages} {061} (\bibinfo {year}
  {2013})},\ \Eprint {http://arxiv.org/abs/1301.1355} {arXiv:1301.1355}
  \BibitemShut {NoStop}%
\bibitem [{\citenamefont {Fodor}\ \emph
  {et~al.}(2012){\natexlab{a}}\citenamefont {Fodor}, \citenamefont {Holland},
  \citenamefont {Kuti}, \citenamefont {Nogradi}, \citenamefont {Schroeder},\
  and\ \citenamefont {Wong}}]{Fodor:2012uw}%
  \BibitemOpen
  \bibfield  {author} {\bibinfo {author} {\bibfnamefont {Z.}~\bibnamefont
  {Fodor}}, \bibinfo {author} {\bibfnamefont {K.}~\bibnamefont {Holland}},
  \bibinfo {author} {\bibfnamefont {J.}~\bibnamefont {Kuti}}, \bibinfo {author}
  {\bibfnamefont {D.}~\bibnamefont {Nogradi}}, \bibinfo {author} {\bibfnamefont
  {C.}~\bibnamefont {Schroeder}}, \ and\ \bibinfo {author} {\bibfnamefont
  {C.~H.}\ \bibnamefont {Wong}},\ }\href@noop {} {\bibfield  {journal}
  {\bibinfo  {journal} {PoS},\ }\textbf {\bibinfo {volume} {Lattice 2012}},\
  \bibinfo {pages} {025} (\bibinfo {year} {2012}{\natexlab{a}})},\ \Eprint
  {http://arxiv.org/abs/1211.3548} {arXiv:1211.3548} \BibitemShut {NoStop}%
\bibitem [{\citenamefont {Fodor}\ \emph
  {et~al.}(2012){\natexlab{b}}\citenamefont {Fodor}, \citenamefont {Holland},
  \citenamefont {Kuti}, \citenamefont {Nogradi}, \citenamefont {Schroeder},\
  and\ \citenamefont {Wong}}]{Fodor:2012et}%
  \BibitemOpen
  \bibfield  {author} {\bibinfo {author} {\bibfnamefont {Z.}~\bibnamefont
  {Fodor}}, \bibinfo {author} {\bibfnamefont {K.}~\bibnamefont {Holland}},
  \bibinfo {author} {\bibfnamefont {J.}~\bibnamefont {Kuti}}, \bibinfo {author}
  {\bibfnamefont {D.}~\bibnamefont {Nogradi}}, \bibinfo {author} {\bibfnamefont
  {C.}~\bibnamefont {Schroeder}}, \ and\ \bibinfo {author} {\bibfnamefont
  {C.~H.}\ \bibnamefont {Wong}},\ }\href@noop {} {\bibfield  {journal}
  {\bibinfo  {journal} {PoS},\ }\textbf {\bibinfo {volume} {Lattice 2012}},\
  \bibinfo {pages} {279} (\bibinfo {year} {2012}{\natexlab{b}})},\ \Eprint
  {http://arxiv.org/abs/1211.4238} {arXiv:1211.4238} \BibitemShut {NoStop}%
\bibitem [{\citenamefont {Aoki}\ \emph {et~al.}(2012)\citenamefont {Aoki},
  \citenamefont {Aoyama}, \citenamefont {Kurachi}, \citenamefont {Maskawa},
  \citenamefont {Nagai}, \citenamefont {Ohki}, \citenamefont {Shibata},
  \citenamefont {Yamawaki},\ and\ \citenamefont {Yamazaki}}]{Aoki:2012eq}%
  \BibitemOpen
  \bibfield  {author} {\bibinfo {author} {\bibfnamefont {Y.}~\bibnamefont
  {Aoki}}, \bibinfo {author} {\bibfnamefont {T.}~\bibnamefont {Aoyama}},
  \bibinfo {author} {\bibfnamefont {M.}~\bibnamefont {Kurachi}}, \bibinfo
  {author} {\bibfnamefont {T.}~\bibnamefont {Maskawa}}, \bibinfo {author}
  {\bibfnamefont {K.-i.}\ \bibnamefont {Nagai}}, \bibinfo {author}
  {\bibfnamefont {H.}~\bibnamefont {Ohki}}, \bibinfo {author} {\bibfnamefont
  {A.}~\bibnamefont {Shibata}}, \bibinfo {author} {\bibfnamefont
  {K.}~\bibnamefont {Yamawaki}}, \ and\ \bibinfo {author} {\bibfnamefont
  {T.}~\bibnamefont {Yamazaki}},\ }\Doi {10.1103/PhysRevD.86.054506} {\bibfield
   {journal} {\bibinfo  {journal} {Phys. Rev.},\ }\textbf {\bibinfo {volume}
  {D86}},\ \bibinfo {pages} {054506} (\bibinfo {year} {2012})},\ \Eprint
  {http://arxiv.org/abs/1207.3060} {arXiv:1207.3060} \BibitemShut {NoStop}%
\bibitem [{\citenamefont {Aoki}\ \emph {et~al.}(2013)\citenamefont {Aoki},
  \citenamefont {Aoyama}, \citenamefont {Kurachi}, \citenamefont {Maskawa},
  \citenamefont {Nagai}, \citenamefont {Ohki}, \citenamefont {Rinaldi},
  \citenamefont {Shibata}, \citenamefont {Yamawaki},\ and\ \citenamefont
  {Yamazaki}}]{Aoki:2013pca}%
  \BibitemOpen
  \bibfield  {author} {\bibinfo {author} {\bibfnamefont {Y.}~\bibnamefont
  {Aoki}}, \bibinfo {author} {\bibfnamefont {T.}~\bibnamefont {Aoyama}},
  \bibinfo {author} {\bibfnamefont {M.}~\bibnamefont {Kurachi}}, \bibinfo
  {author} {\bibfnamefont {T.}~\bibnamefont {Maskawa}}, \bibinfo {author}
  {\bibfnamefont {K.-i.}\ \bibnamefont {Nagai}}, \bibinfo {author}
  {\bibfnamefont {H.}~\bibnamefont {Ohki}}, \bibinfo {author} {\bibfnamefont
  {E.}~\bibnamefont {Rinaldi}}, \bibinfo {author} {\bibfnamefont
  {A.}~\bibnamefont {Shibata}}, \bibinfo {author} {\bibfnamefont
  {K.}~\bibnamefont {Yamawaki}}, \ and\ \bibinfo {author} {\bibfnamefont
  {T.}~\bibnamefont {Yamazaki}},\ }\href@noop {} { (\bibinfo {year} {2013})},\
  \Eprint {http://arxiv.org/abs/1302.4577} {arXiv:1302.4577} \BibitemShut
  {NoStop}%
\bibitem [{\citenamefont {Itou}(2013)}]{Itou:2012qn}%
  \BibitemOpen
  \bibfield  {author} {\bibinfo {author} {\bibfnamefont {E.}~\bibnamefont
  {Itou}},\ }\Doi {10.1093/ptep/ptt053} {\bibfield  {journal} {\bibinfo
  {journal} {PTEP},\ }\textbf {\bibinfo {volume} {2013}},\ \bibinfo {pages}
  {083B01} (\bibinfo {year} {2013})},\ \Eprint {http://arxiv.org/abs/1212.1353}
  {arXiv:1212.1353} \BibitemShut {NoStop}%
\bibitem [{\citenamefont {Lin}\ \emph {et~al.}(2012)\citenamefont {Lin},
  \citenamefont {Ogawa}, \citenamefont {Ohki},\ and\ \citenamefont
  {Shintani}}]{Lin:2012iw}%
  \BibitemOpen
  \bibfield  {author} {\bibinfo {author} {\bibfnamefont {C.-J.~D.}\
  \bibnamefont {Lin}}, \bibinfo {author} {\bibfnamefont {K.}~\bibnamefont
  {Ogawa}}, \bibinfo {author} {\bibfnamefont {H.}~\bibnamefont {Ohki}}, \ and\
  \bibinfo {author} {\bibfnamefont {E.}~\bibnamefont {Shintani}},\ }\Doi
  {10.1007/JHEP08(2012)096} {\bibfield  {journal} {\bibinfo  {journal} {JHEP},\
  }\textbf {\bibinfo {volume} {1208}},\ \bibinfo {pages} {096} (\bibinfo {year}
  {2012})},\ \Eprint {http://arxiv.org/abs/1205.6076} {arXiv:1205.6076}
  \BibitemShut {NoStop}%
\bibitem [{\citenamefont {Jin}\ and\ \citenamefont
  {Mawhinney}(2012)}]{Jin:2012dw}%
  \BibitemOpen
  \bibfield  {author} {\bibinfo {author} {\bibfnamefont {X.-Y.}\ \bibnamefont
  {Jin}}\ and\ \bibinfo {author} {\bibfnamefont {R.~D.}\ \bibnamefont
  {Mawhinney}},\ }\href@noop {} {\bibfield  {journal} {\bibinfo  {journal}
  {PoS},\ }\textbf {\bibinfo {volume} {Lattice 2011}},\ \bibinfo {pages} {066}
  (\bibinfo {year} {2012})},\ \Eprint {http://arxiv.org/abs/1203.5855}
  {arXiv:1203.5855} \BibitemShut {NoStop}%
\bibitem [{\citenamefont {Hasenfratz}\ \emph
  {et~al.}(2013){\natexlab{a}}\citenamefont {Hasenfratz}, \citenamefont
  {Cheng}, \citenamefont {Petropoulos},\ and\ \citenamefont
  {Schaich}}]{Hasenfratz:2013uha}%
  \BibitemOpen
  \bibfield  {author} {\bibinfo {author} {\bibfnamefont {A.}~\bibnamefont
  {Hasenfratz}}, \bibinfo {author} {\bibfnamefont {A.}~\bibnamefont {Cheng}},
  \bibinfo {author} {\bibfnamefont {G.}~\bibnamefont {Petropoulos}}, \ and\
  \bibinfo {author} {\bibfnamefont {D.}~\bibnamefont {Schaich}},\ }\href@noop
  {} { (\bibinfo {year} {2013}{\natexlab{a}})},\ \Eprint
  {http://arxiv.org/abs/1303.7129} {arXiv:1303.7129} \BibitemShut {NoStop}%
\bibitem [{\citenamefont {Hasenfratz}\ \emph
  {et~al.}(2013){\natexlab{b}}\citenamefont {Hasenfratz}, \citenamefont
  {Cheng}, \citenamefont {Petropoulos},\ and\ \citenamefont
  {Schaich}}]{Hasenfratz:2013eka}%
  \BibitemOpen
  \bibfield  {author} {\bibinfo {author} {\bibfnamefont {A.}~\bibnamefont
  {Hasenfratz}}, \bibinfo {author} {\bibfnamefont {A.}~\bibnamefont {Cheng}},
  \bibinfo {author} {\bibfnamefont {G.}~\bibnamefont {Petropoulos}}, \ and\
  \bibinfo {author} {\bibfnamefont {D.}~\bibnamefont {Schaich}},\ }\href@noop
  {} {\bibfield  {journal} {\bibinfo  {journal} {PoS},\ }\textbf {\bibinfo
  {volume} {LATTICE 2013}},\ \bibinfo {pages} {075} (\bibinfo {year}
  {2013}{\natexlab{b}})},\ \Eprint {http://arxiv.org/abs/1310.1124}
  {arXiv:1310.1124} \BibitemShut {NoStop}%
\bibitem [{\citenamefont {DeGrand}\ and\ \citenamefont
  {Hasenfratz}(2009)}]{DeGrand:2009mt}%
  \BibitemOpen
  \bibfield  {author} {\bibinfo {author} {\bibfnamefont {T.}~\bibnamefont
  {DeGrand}}\ and\ \bibinfo {author} {\bibfnamefont {A.}~\bibnamefont
  {Hasenfratz}},\ }\Doi {10.1103/PhysRevD.80.034506} {\bibfield  {journal}
  {\bibinfo  {journal} {Phys. Rev.},\ }\textbf {\bibinfo {volume} {D80}},\
  \bibinfo {pages} {034506} (\bibinfo {year} {2009})},\ \Eprint
  {http://arxiv.org/abs/0906.1976} {arXiv:0906.1976} \BibitemShut {NoStop}%
\bibitem [{\citenamefont {Del~Debbio}\ and\ \citenamefont
  {Zwicky}(2010)}]{DelDebbio:2010ze}%
  \BibitemOpen
  \bibfield  {author} {\bibinfo {author} {\bibfnamefont {L.}~\bibnamefont
  {Del~Debbio}}\ and\ \bibinfo {author} {\bibfnamefont {R.}~\bibnamefont
  {Zwicky}},\ }\Doi {10.1103/PhysRevD.82.014502} {\bibfield  {journal}
  {\bibinfo  {journal} {Phys. Rev.},\ }\textbf {\bibinfo {volume} {D82}},\
  \bibinfo {pages} {014502} (\bibinfo {year} {2010})},\ \Eprint
  {http://arxiv.org/abs/1005.2371} {arXiv:1005.2371} \BibitemShut {NoStop}%
\bibitem [{\citenamefont {Hasenbusch}(1999)}]{Hasenbusch:1999mw}%
  \BibitemOpen
  \bibfield  {author} {\bibinfo {author} {\bibfnamefont {M.}~\bibnamefont
  {Hasenbusch}},\ }\Doi {10.1088/0305-4470/32/26/304} {\bibfield  {journal}
  {\bibinfo  {journal} {J.Phys.},\ }\textbf {\bibinfo {volume} {A32}},\
  \bibinfo {pages} {4851} (\bibinfo {year} {1999})},\ \Eprint
  {http://arxiv.org/abs/hep-lat/9902026} {arXiv:hep-lat/9902026 [hep-lat]}
  \BibitemShut {NoStop}%
\bibitem [{\citenamefont {Hasenbusch}(2010)}]{Hasenbusch:2011yya}%
  \BibitemOpen
  \bibfield  {author} {\bibinfo {author} {\bibfnamefont {M.}~\bibnamefont
  {Hasenbusch}},\ }\Doi {10.1103/PhysRevB.82.174433} {\bibfield  {journal}
  {\bibinfo  {journal} {Phys.Rev.},\ }\textbf {\bibinfo {volume} {B82}},\
  \bibinfo {pages} {174433} (\bibinfo {year} {2010})}\BibitemShut {NoStop}%
\bibitem [{\citenamefont {Del~Debbio}\ and\ \citenamefont
  {Zwicky}(2013)}]{DelDebbio:2013qta}%
  \BibitemOpen
  \bibfield  {author} {\bibinfo {author} {\bibfnamefont {L.}~\bibnamefont
  {Del~Debbio}}\ and\ \bibinfo {author} {\bibfnamefont {R.}~\bibnamefont
  {Zwicky}},\ }\href@noop {} { (\bibinfo {year} {2013})},\ \Eprint
  {http://arxiv.org/abs/1306.4038} {arXiv:1306.4038} \BibitemShut {NoStop}%
\bibitem [{\citenamefont {Lepage}\ \emph {et~al.}(2002)\citenamefont {Lepage},
  \citenamefont {Clark}, \citenamefont {Davies}, \citenamefont {Hornbostel},
  \citenamefont {Mackenzie} \emph {et~al.}}]{Lepage:2001ym}%
  \BibitemOpen
  \bibfield  {author} {\bibinfo {author} {\bibfnamefont {G.}~\bibnamefont
  {Lepage}}, \bibinfo {author} {\bibfnamefont {B.}~\bibnamefont {Clark}},
  \bibinfo {author} {\bibfnamefont {C.}~\bibnamefont {Davies}}, \bibinfo
  {author} {\bibfnamefont {K.}~\bibnamefont {Hornbostel}}, \bibinfo {author}
  {\bibfnamefont {P.}~\bibnamefont {Mackenzie}},  \emph {et~al.},\ }\Doi
  {10.1016/S0920-5632(01)01638-3} {\bibfield  {journal} {\bibinfo  {journal}
  {Nucl.Phys.Proc.Suppl.},\ }\textbf {\bibinfo {volume} {106}},\ \bibinfo
  {pages} {12} (\bibinfo {year} {2002})},\ \Eprint
  {http://arxiv.org/abs/hep-lat/0110175} {arXiv:hep-lat/0110175 [hep-lat]}
  \BibitemShut {NoStop}%
\bibitem [{\citenamefont {Hornbostel}\ \emph {et~al.}(2012)\citenamefont
  {Hornbostel}, \citenamefont {Lepage}, \citenamefont {Davies}, \citenamefont
  {Dowdall}, \citenamefont {Na} \emph {et~al.}}]{Hornbostel:2011hu}%
  \BibitemOpen
  \bibfield  {author} {\bibinfo {author} {\bibfnamefont {K.}~\bibnamefont
  {Hornbostel}}, \bibinfo {author} {\bibfnamefont {G.}~\bibnamefont {Lepage}},
  \bibinfo {author} {\bibfnamefont {C.}~\bibnamefont {Davies}}, \bibinfo
  {author} {\bibfnamefont {R.}~\bibnamefont {Dowdall}}, \bibinfo {author}
  {\bibfnamefont {H.}~\bibnamefont {Na}},  \emph {et~al.},\ }\Doi
  {10.1103/PhysRevD.85.031504} {\bibfield  {journal} {\bibinfo  {journal}
  {Phys.Rev.},\ }\textbf {\bibinfo {volume} {D85}},\ \bibinfo {pages} {031504}
  (\bibinfo {year} {2012})},\ \Eprint {http://arxiv.org/abs/1111.1363}
  {arXiv:1111.1363 [hep-lat]} \BibitemShut {NoStop}%
\bibitem [{Note1()}]{Note1}%
  \BibitemOpen
  \bibinfo {note} {\protect \texttt
  {https://github.com/gplepage/corrfitter}}\BibitemShut {NoStop}%
\bibitem [{\citenamefont {Ishikawa}\ \emph {et~al.}(2013)\citenamefont
  {Ishikawa}, \citenamefont {Iwasaki}, \citenamefont {Nakayama},\ and\
  \citenamefont {Yoshie}}]{Ishikawa:2013tua}%
  \BibitemOpen
  \bibfield  {author} {\bibinfo {author} {\bibfnamefont {K.~I.}\ \bibnamefont
  {Ishikawa}}, \bibinfo {author} {\bibfnamefont {Y.}~\bibnamefont {Iwasaki}},
  \bibinfo {author} {\bibfnamefont {Y.}~\bibnamefont {Nakayama}}, \ and\
  \bibinfo {author} {\bibfnamefont {T.}~\bibnamefont {Yoshie}},\ }\href@noop {}
  { (\bibinfo {year} {2013})},\ \Eprint {http://arxiv.org/abs/1310.5049}
  {arXiv:1310.5049 [hep-lat]} \BibitemShut {NoStop}%
\bibitem [{Note2()}]{Note2}%
  \BibitemOpen
  \bibinfo {note} {\protect \texttt {http://www.physics.utah.edu/$\sim
  $detar/milc/}}\BibitemShut {NoStop}%
\end{thebibliography}%
  \bibliographystyle{apsrev}
}
\end{document}